# Diamond of Thought: A Design Thinking-Based Framework for LLMs in Wearable Design

Qiyang Miao

College of Design and Innovation, Tongji University, 200092 Shanghai, China, mqy0436@126.com

Jiang Xu*

College of Design and Innovation, Tongji University, 200092 Shanghai, China, Xuj@tongji.edu.com

Zhihao Song

College of Design and Innovation, Tongji University, 200092 Shanghai, China, wittywizardsong@outlook.com

Chengrui Wang

Faculty of Arts and Humanities, Coventry University, CV1 5FB United Kingdom, wangc64@uni.coventry.ac.uk

Yu Cui

College of Design and Innovation, Tongji University,200092 Shanghai, China, cuiyu_1997@163.com

**ABSTRACT:** Wearable design is an interdisciplinary field that balances technological innovation, human factors, and human-computer interactions. Despite contributions from various disciplines, many projects lack stable interdisciplinary teams, which often leads to design failures. Large language models (LLMs) integrate diverse information and generate innovative solutions, making them a valuable tool for enhancing design processes. Thus, we have explored the use of LLMs in wearable design by combining design-thinking principles with LLM capabilities. We have developed the "Diamond of Thought" framework and analysed 1,603 prototypes and 1,129 products from a body-centric perspective to create a comprehensive database. We employed retrieval-augmented generation to input database details into the LLMs, ensuring applicability to wearable design challenges and integration of embodied cognition into the process. Our LLM-based methodology for wearables has been experimentally validated, demonstrating the potential of LLMs for the advancement of design practices. This study offers new tools and methods for future wearable designs.

## Conflict of Interest

The authors declare no conflict of interests

___________________________

*Corresponding author

# 1 INTRODUCTION

The emerging field of wearable design is recognised as an interdisciplinary domain, bridging engineering, human factors, and human-computer interactions (HCI) [1]. Although each discipline contributes valuable knowledge, not all wearable computing projects involve a stable interdisciplinary team [2]. This lack of cross-disciplinary expertise often leads to the failure of wearable designs.

Advancements in technology and increases in the processing capacity of artificial intelligence have allowed large language models (LLMs) to achieve notable success in natural language processing and have engendered innovations in other domains. In the design field, LLMs can understand design problems, integrate interdisciplinary knowledge, and generate textual solutions. They excel at synthesising information from diverse sources and disciplines, offering a promising approach for integrating interdisciplinary knowledge into the design process. Many studies have begun to explore the application of LLMs in design [3, 4, 5] by combining these models with generative design techniques to facilitate the creation of more innovative solutions.

We propose the Diamond of Thought (DoT) as a framework for actively solving concept design tasks with LLMs, focusing specifically on its potential applications within the domain of wearable design. We posit that this method could evolve LLMs into a sophisticated conceptual design tool with superior problem-solving and information integration capabilities. Design-thinking principles and standard methodologies for addressing design challenges serve as the foundation of our work. We aligned the divergent and convergent structures required at each stage of the design task to complete the generation and selection of nodes at various levels. Concurrently, adopting a body-centric perspective, we collected and analysed 1,603 wearable prototypes and 1,129 wearable products and coded them by function, behaviour, and structure. This comprehensive analysis resulted in an extensive wearable technology database [6] encompassing a broad and interdisciplinary range of knowledge related to technology, functionality, and the human body from past wearable design cases. We employed retrieval-augmented generation (RAG) to input database details into the LLMs in steps, affording them the ability to manage wearable design challenges and incorporate embodied cognition into the design process.

To assess the effectiveness of the proposed methods, we conducted a conceptual design experiment comparing LLMs with human designers to thoroughly evaluate their abilities, strengths, and overall performance. The results indicate that our newly proposed DoT framework, combined with our overall methods, allows LLMs to effectively integrate past cases and knowledge to generate rational and comprehensive wearable design solutions.

# 2 RELATED WORK

## 2.1 Wearable design

Wearable devices are defined as any form of miniature electronic or sensor technology worn on the body [7]. In recent years, the use of wearable technology has significantly increased. Devices such as smartphones, activity trackers, and smart watches are widely used and seem to have become almost inseparable from the human body [8]. The field of wearable technologies is recognised as an interdisciplinary domain, with existing literature spanning many disciplines, including engineering, social sciences, and health sciences [9]. Previous research suggests that, if designed effectively, wearable devices can leverage cultural, social, and sociological factors, as well as the diverse meanings of the body, in conjunction with the features of smart and programmable elements [10]. Joseph et al. proposed that wearable device design should be approached from the perspectives of fashion theory, somatics, and digital materialities [1]. However, interdisciplinary expertise is often lacking in the current design process [2]. Collaboration between disciplines is challenging, and designers with a single background may lack the comprehensive interdisciplinary knowledge required for effective integration [11]. Support from diverse experts is essential to enhance problem-solving success.

Moreover, the discourse on wearable technologies and informed use of HCI design methods have predominantly followed scientific paradigms [1]. Such approaches analysed wearable devices from a third-person perspective, viewing them as autonomous material objects placed on the body, containers for body quantification data, or tools for achieving specific human objectives [12]. This analytical perspective, grounded in mind-body dualism, tends to overlook the unique characteristics of wearable devices. Thus, conventional approaches may yield devices that are not effectively integrated with the human body, consequently increasing the usability threshold [6].

## 2.2 LLMs

Advancements in LLMs have illuminated the path towards artificial general intelligence, with notable examples including generative pre-trained transformers (GPTs) [13, 14], LLaMA4 [15], and Geminis [16]. Because they have demonstrated capabilities close to the human level on tasks of understanding natural language, text generation, and reasoning, LLMs have extensive applications in various fields, such as medicine [17], mathematics [18], and chemistry [19]. Despite the significant advancements in LLMs, several challenges remain. A notable example is hallucinations, whereby LLMs generate incorrect or fabricated information that can mislead users. Researchers have proposed various strategies to enhance LLM capabilities and mitigate hallucinations, including practical prompt engineering, identity setting, lavish praise, few-shot prompting, and RAG [20, 21, 22].

Wei et al. introduced a problem-solving approach called the Chain of Thought (CoT) that significantly enhanced the problem-solving capabilities of LLMs [23]. Jin et al. further demonstrated that complex tasks benefited more from extended reasoning sequences [24]. Wang et al. introduced self-consistency with CoT; their approach was proven to effectively select the best solution following a consistency evaluation of multiple CoT outcomes [25]. Yao et al. revisited the human logical reasoning process, highlighting that human reasoning and decision-making for complex problems are not linear. They revisited Newell's early topology of the human problem-solving process [26] and proposed the Tree of Thoughts (ToT) as a method of searching within the mental space in the form of a tree [27]. Maciej et al. improved data structures by introducing a problem-solving method that is closer to human thinking, i.e., the Graph of Thoughts [28]. Ding et al. proposed addressing the limitations of the "Penrose Triangle" by applying the concept of Everything of Thought to co-optimise performance, efficiency, and adaptability [29]. This body of work collectively underscores a significant trend in the evolution of problem-solving methodologies in LLMs, driving them towards models that more closely mimic human cognitive processes and adapt more robustly to complex real-world tasks.

### 2.3  LLM-based generative conceptual design

LLMs have numerous benefits that render them exceedingly effective in various fields. LLM-based designs have emerged as a novel paradigm that has garnered a considerable amount of interest. With access to pre-collected design knowledge [30], LLMs can define designs based on the inputs of problem-specific keywords and leverage previously learned diverse design styles, creativity, and information from various disciplines and fields to generate potential design solutions. This approach ensures that LLMs establish a thorough understanding of existing design cases and solutions while drawing extensively from the fields of engineering, cognitive sciences, and creative arts. Consequently, their integration into conceptual design has become particularly valuable, enhancing the innovative potential of design solutions.

Zhu and Luo's innovative exploration of GPTs for verbal design concepts utilised customised datasets to test the capacity of GPT models for problem- and analogy-driven reasoning. The findings revealed that GPT-2 and GPT-3 could effectively harness learned design styles and cross-domain information to create novel design solutions [3]. Building on this, Ma et al. investigated the ability of LLMs to generate conceptual design solutions by comparing them with those derived via crowdsourcing methods [4]. They observed that prompt engineering and few-shot learning significantly enhanced the design outputs, highlighting that LLMs, particularly when optimised with these techniques, offer substantial support during the conceptual design phase. However, a major challenge in LLM design is their opacity and limited controllability, which often prevent them from producing design concepts that precisely match specific problems and requirements. To address this challenge, Wang et al. introduced an innovative methodology that integrates LLMs with a function-behaviour-structure (FBS) model to facilitate generative conceptual design [5]. The results demonstrated that the designs generated via this task-decomposition method were more creative and closely aligned with the design requirements.

In conclusion, LLM-based design has considerable potential for innovation. Techniques such as prompt engineering, the fine-tuning of pre-trained models, and task decomposition are instrumental in advancing LLM-based design technology, as they open new avenues for research and application in the field of design innovation.

## 3  DIAMOND OF THOUGHT

### 3.1  Construction of the LLM process framework for concept design generation

Problem-solving within the design-thinking paradigm is inherently complex, dynamic, and iterative. Designers must flexibly adjust their directions and navigate through numerous divergences and convergences throughout the process.

This necessitates a departure from conventional thought patterns, embracing divergent thinking to generate a multitude of creative ideas and solutions and subsequently integrating selected directions to adapt to constantly changing needs and conditions. To effectively harness LLMs for conceptual design tasks, it is essential to first define the characteristics and overall processes of this complex cognitive model and then systematically structure it to fully exploit the advantages of LLMs.

Building on previous design–thinking models, we developed the DoT framework, an enhanced conceptual design task–focused framework grounded in the thinking tree methodology. This framework was designed to emulate the fundamental design–thinking characteristics and problem-solving flows of designers. The DoT framework allows LLMs to deconstruct design tasks and address them sequentially, drawing on design knowledge to fulfil the tasks. The diamond metaphor represents the creation and filtering of nodes, capturing the essence of design thinking through the interplay of divergence and convergence to ensure that the iterative process not only generates new ideas, it also keeps them relevant and targeted.

As illustrated in Figure 1, our DoT framework is characterised by a multistep process with alternating divergence and convergence based on a database and user input information framework. Waidelich et al. conducted a comprehensive analysis of 35 design–thinking models, tallying the frequency of different stages across all models: "Ideation/Ideate" appeared 30 times, "Prototype/Prototyping" 22 times, "Test/Testing" 20 times, "Define" 17 times, and "Understanding/Understand" 14 times [31]. Accordingly, we structured the process of using LLMs to solve design tasks as follows:

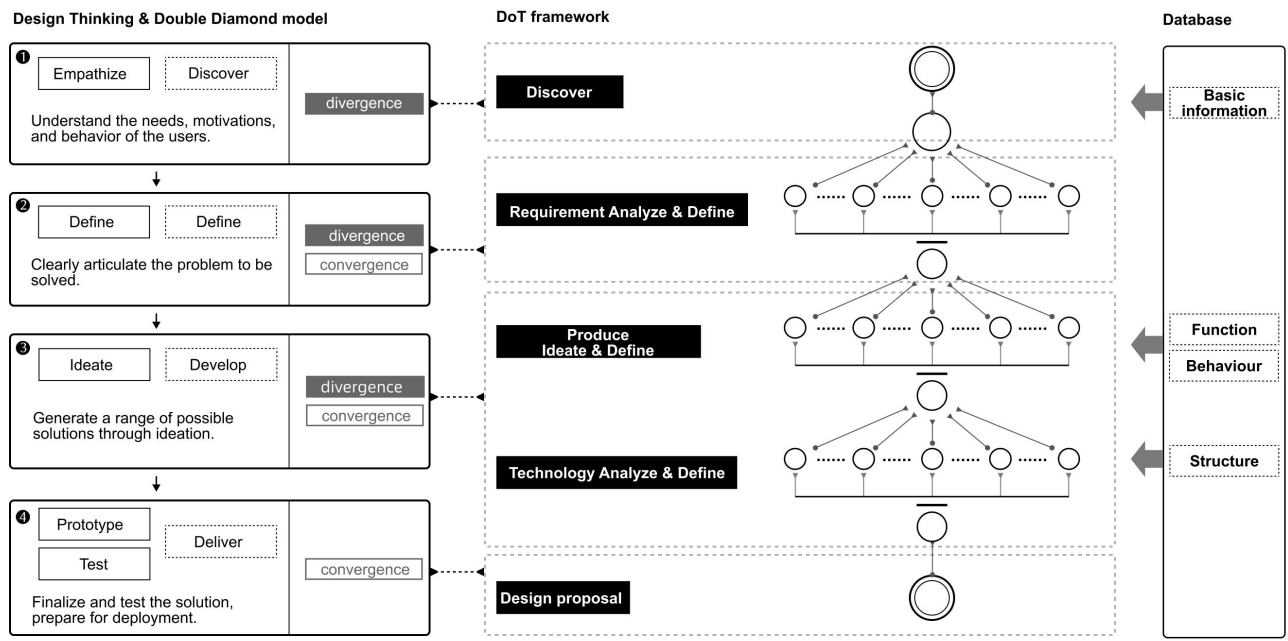

**Fig.1: DOT framework**

1. Discovery: The design process begins by exploring the contextual background of the design task whereby the LLM engages in both active and passive information collection.

2. Requirement Analysis and Definition: This stage identifies and prioritises potential needs from the accumulated information, filtering them according to their significance and the designers' inclinations.

3. Produce, Ideate, and Define: Leveraging previously acquired information and knowledge, this phase involves divergent thinking regarding a product's functionalities, followed by a process of comparison, selection, and integration to finalise the product's definition.

4. Technology Analysis and Definition: This phase involves numerous integrations using the given design database cases and inherent knowledge of LLMs to devise concrete technical strategies, followed by comparison, filtering, and amalgamation of these strategies to solidify the ultimate technical strategy.

Unlike previous models, we did not allocate divergent or convergent thinking to specific stages. Instead, we maintain that several stages require a divergence–convergence cycle that serves as a strategy for self–consistency, a method already validated for its efficacy. This approach is applied throughout the entire design process.

### 3.2 Application of design case databases

Human designers often rely on external knowledge and past designs to complete their design tasks. Relevant design cases offer valuable insight into the background information, current solutions, and potential future pathways for addressing specific design problems. They serve as a crucial source of inspiration for the innovation and exploration of new design opportunities. These cases also provide established strategies for solving design problems and demonstrate how previous designs have successfully integrated cross–disciplinary knowledge, such as technical applications, material selection, and structural principles, to address complex challenges. This allows designers to apply similar strategies and methods in their current practices. Additionally, these cases inspire designers to explore untapped opportunities, helping them to identify unmet needs and potential pain points.

Our previous work integrated data from prominent databases [6], including Web of Science, Scopus, IEEE, DBLP, and NCBI, as well as information from globally recognised sources, such as Wearable Technologies, Fabric of Digital Life, YouTube, Apple, SONY, and Fitbit. This effort resulted in the curation of a comprehensive dataset comprising 1,603 wearable device prototypes and 1,129 market products, as shown in Figure 2. The human body serves as the platform for wearable devices and plays a pivotal role in their development and design. Thus, we adopted an embodied aspect and developed an encoding method for wearable device data based on the embodied FBS ontology theory. Specifically, the functionality of wearable devices is dependent on the requirements necessitated by the embodied existence of the body in the world. The perception–motor coupling system within the body schema facilitates an interactive channel through which wearable objects can interface with the body, whereas the physical characteristics inherent to the body provide the foundational space for the distribution of wearable objects. Consequently, the database focuses on the specific functions of wearable devices at the functional level, analyses the perception and action modes of the interaction between the body and wearable devices at the behavioural level, and considers the technical elements, on–body locations, and modes of wearing at the structural level. This database provides a foundation for the subsequent production of wearable designs through the integration of multiple perspectives and embodied cognition.

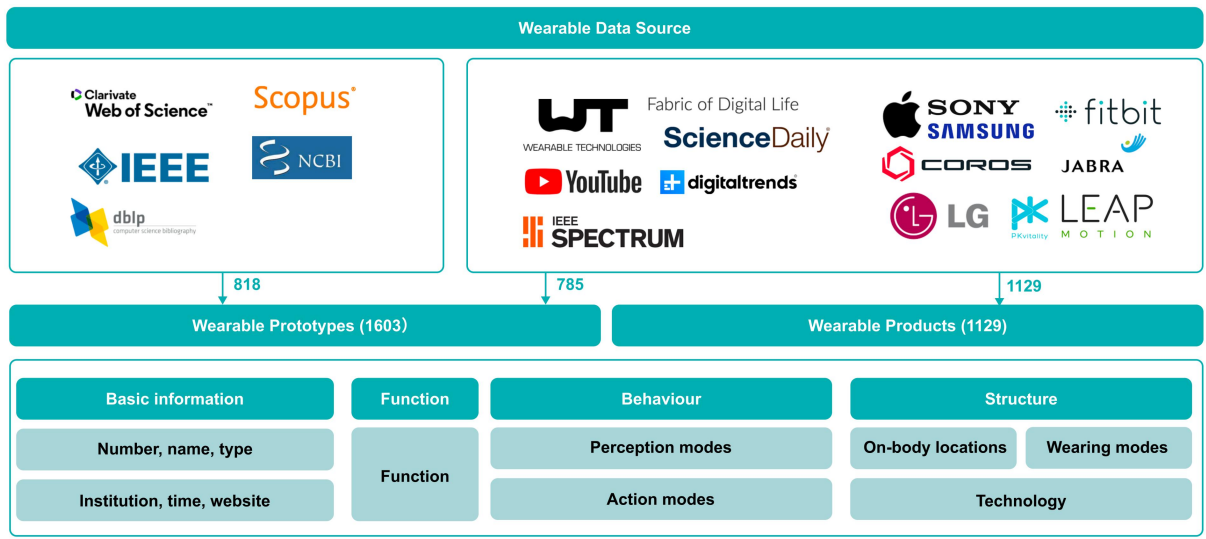

Fig.2: Wearable database structure (Xu, Hao, Sun, Qu, Mo, Wang, & Xu, 2023)

Previous research has shown that providing appropriate prompts to LLMs can inspire and guide models to generate higher quality and more innovative design concepts [5]. Consequently, at each design stage, we input different categories of information from the database into the LLMs in the form of prompts, as illustrated in Figure 3. Specifically, in the discovery phase, we initially expose the LLM to case data to understand the broad context of the design task, thereby enhancing the understanding and definitions of the design problems and requirements. Based on the background information provided, the LLM identifies possible categories of target devices, finds the best products in the wearable knowledge vector database, and includes these products and related data in the design process as benchmarks for future design tasks. In the Produce, Ideate, and Define phase, we input information related to the functional and behavioural aspects of database cases into the LLM to facilitate deeper insight into the existing design case solution strategies. Additionally, in the subsequent Technology Analysis and Definition stage, we input structural information into the LLM to ensure that it generates more reasonable technical solutions based on the existing FBS matching relationships.

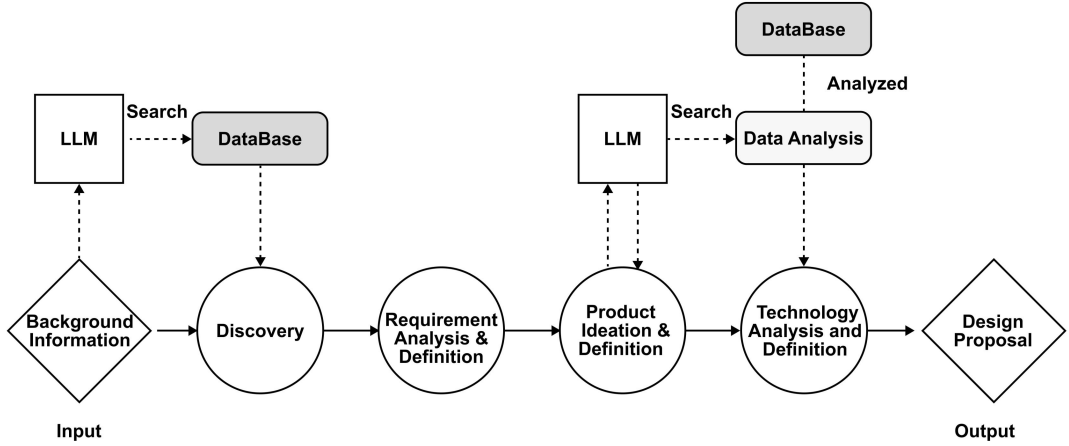

Fig.3: Use of databases for concept generation

### 3.3 Node Divergence and Convergence

In this section, we provide further details on the node divergence and convergence strategies of the DOT framework.

**Thought divergence strategy.** When a new design stage is reached, the interim results from the previous phase are analysed and used as a foundation for subsequent design solution development. We aimed to closely simulate the real-world problem-solving processes of designers and maximally incorporate divergent thinking. During the several rounds of divergent thinking, we focus on several key points. First, designers' thought processes are continuous, as they consider design problems holistically, rather than addressing each step in isolation. Additionally, designers do not limit their problem solving to their own knowledge, instead exploring external information and multidisciplinary insights to optimise design outcomes. Thus, prompts were carefully crafted to accurately simulate the design process.

The initial information and previous design output are respectively represented by $x$ and $y$, and the current layer design output (i.e., the state of the tree) is represented by

$$s = [x, y_1 ... y_i] \quad (1)$$

From this, we can introduce our design idea generator as follows:

$$y_{i+1}^{1...k} = [y_{i+1}^1 ... y_{i+1}^j] \, p_\theta \, (y_{i+1}^{1...j} \vee s) \quad (2)$$

We chose a top-down continuous structure for the thinking tree, allowing extensive divergence in each layer to simulate the divergent phase of design thinking. In conceptual design solutions, frequently recurring solutions may represent general conventional approaches to solving design problems, often with high feasibility. In contrast, less frequent solutions can be more creative and inspire innovative approaches to the problem. Within this tree structure, LLMs often generate numerous repeated nodes under synonymous prompts. Although this repetition may limit the cognitive space, it also helps the LLM to assess the importance of different solutions. Thus, meaningful repetition was retained during the divergence phase.

**Thought convergence strategy.** In addition to extensive divergence, optimising the design plan requires effective convergence to reach a phased consensus. Uncertainty in design direction and excessive divergence can significantly limit the design problem solving efficiency. Some design methodologies view the core of innovative problem solving as finding optimal solutions among multiple conflicting needs. Therefore, during the node–generation phase, we allow the LLM to generate repetitive ideas within a limited cognitive space. Subsequently, in the convergence phase, we comprehensively evaluate the divergent results to formulate the next design strategy.

The tree state before node converge can be expressed as

$$s = [x, y_1...y_i] \quad (3)$$

The output of the thought node generation phase is as follows:

$$s_G = [y_{i+1}^1...y_{i+1}^j] \quad (4)$$

In this layer, the thought process should incorporate the nodes generated in the previous step and all of the information established within the previous layers. However, changes in these nodes are not applied for selection or comparison, but rather a synthesis based on existing information. Thus, we introduce a design–node synthesiser:

$$y_{i+1} p_\theta (y_{i+1} \vee s, s_G) \quad (5)$$

The process of staged convergence in design involves eliminating redundancies or merging ideas. This requires a comprehensive understanding of the design problem, integrating multiple perspectives, and balancing potential conflicts to develop an optimal strategy. During this process, many potential solutions and decisions are not entirely clear. These are gradually revealed and utilised through progressive exploration and optimisation processes. Therefore, the proposed node–filtering strategy is based on a comprehensive self-consistency assessment. None of the possible solutions ($p$) is fully explicit in the reasoning process. Instead, through prompts and logical planning, the LLM can capture potential and optimal solutions by using contextual clues and logical relationships.

**Prompt strategy.** Prompts serve as a fundamental pathway for interaction with LLMs and play a crucial role in the generation and aggregation of nodes. At specific stages of the tasks, we provide the LLM with prompts of similar structures to guide their simulation of specific design steps. Additionally, we aimed to enhance the maintainability and interchangeability of the prompts.

We designed specific design methods $m$, example prompts $e$, and frequencies $n$, employing the function () to call various parameter lists across different steps.

$$\pi: m \times e \times n \rightarrow Params \quad (6)$$

Consequently, the prompt aggregator has the following comprehensive framework:

$$P(step) = P(\pi(step)) \quad (7)$$

### 3.4 Searching Algorithm

According to Cross, design experts typically employ top-down, breadth-first strategies and explicit problem decomposition techniques [32]. Inspired by these practices, our algorithm adopts breadth-first search (BFS) as the traversal method for the DOT optimisation, applying a structure similar to that of ToT-BFS algorithms [27]. Specifically, nodes are generated by applying the aforementioned generation strategy during level-order traversal, pruning is achieved under the selection strategy, and the prompt generation method enhances the implementation of these strategies. It is important to note that this algorithm primarily describes the divergent and convergent phases of the DOT, which only occur in the iterative stages of the design process, i.e., requirement analysis, functional definition, and technical solution generation.

---

DoT-BFS Algorithm $(b, p_\theta, G, P, S, k, d_{max})$

---

Requirement: Input design background $b$, LLM $p_\theta$, design ideas generator $G(\ )$, design ideas selector $S(\ )$, Prompt generator $P(\ )$, step limit $d_{max}$, design solution $D$, database $DB$, phased results $R$

$D \leftarrow \emptyset$

While $d \leq d_{max}$ do

$\quad R'_{step} \leftarrow \{[r, z] \vee r \in R_{step-1}, z \in G(p_\theta, P(R_{step-1}, b, k, step, DB))\}$

$\quad R_{step} \leftarrow S(p_\theta, P(R'_{step}, b, k, step, DB))$

$\quad D \leftarrow D \cup R_{step}$

$\quad d \leftarrow d + 1$

end while

return $D$

---

(8)

### 3.5 DOT framework-based LLM task processing procedure

Taking the task of designing a "smart comfort anti-epidemic mask" as an example, we demonstrate the design task processing procedure of the LLM's DOT framework, called the GPT-4 API, with the temperature parameter set to 0.7.

This process is illustrated in Figure 4.

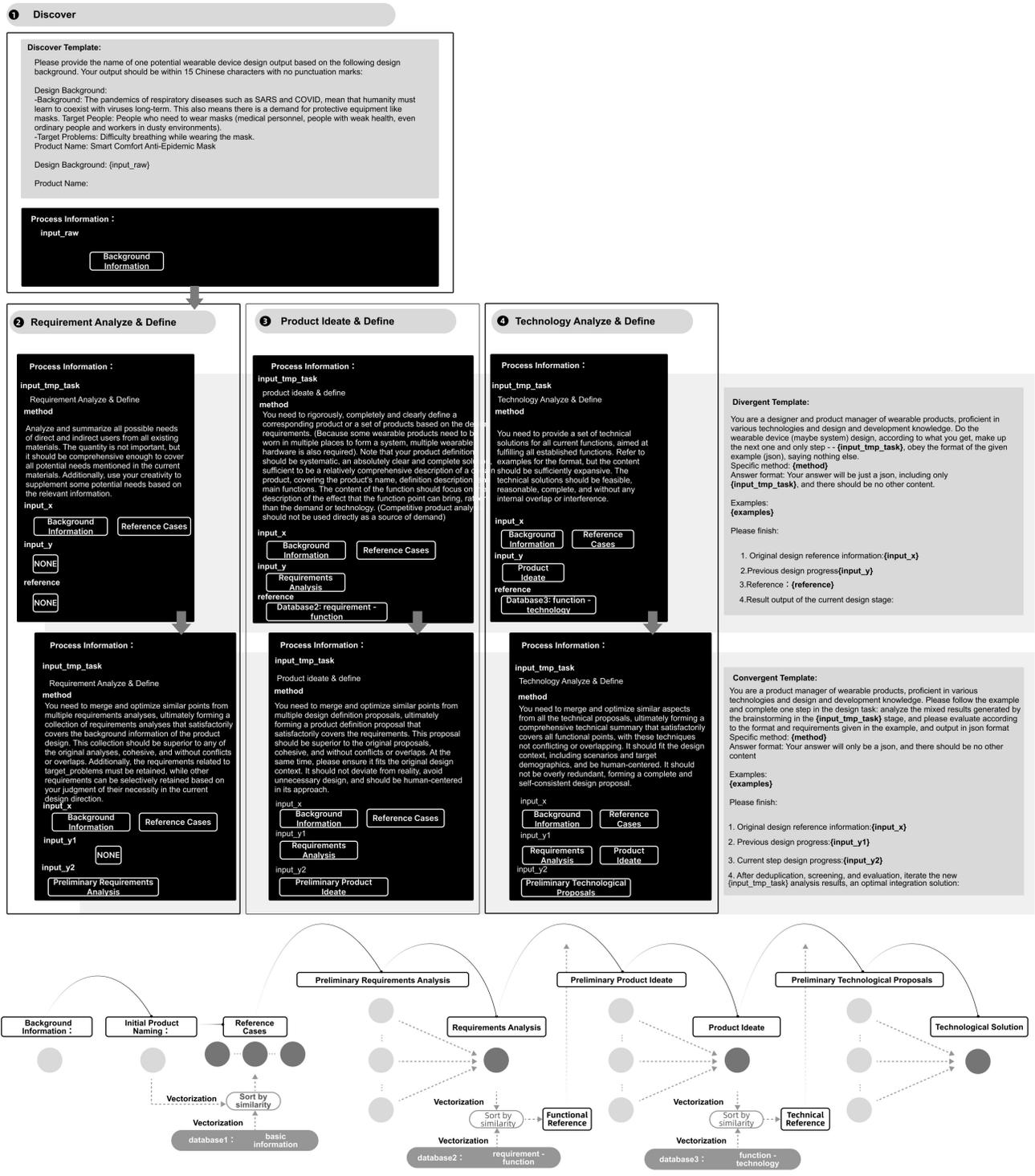

Fig.4: DOT framework-based LLM task processing procedure

## 4 EXPERIMENT AND EVALUATION

We designed a comparative experiment to verify the effectiveness of enhancing the ability of the LLM to solve design problems within the DOT framework.

### 4.1 Participants

As shown in Figure 5, in addition to the DOT group, we setup three control groups to perform conceptual design tasks to test its effectiveness. It is worth mentioning that all subjects used the GPT-4 API with the temperature parameter set at 0.7. The first group was the IO group included a zero-shot GPT-4 model , which was provided explicit instructions for accomplishing the design goal without any limitations on the content or format of its output. The second subject group was the CoT group, which comprised the GPT-4 model with the CoT framework. The one-shot format prompts allowed the LLM responses to incorporate various stages into the design process. Additionally, the DOT group adopted the proposed framework. This framework allows the LLM to think step by step, absorb new knowledge, diverge and converge thoughts in multiple rounds of dialogue, and arrive at a final overall design solution. Lastly, a group consisting of human designers was supplied with a zero-shot design requirement document, allowing them to independently structure the design process and produce design solutions based on their expertise and understanding. In this study, we collected data through questionnaires and face-to-face interviews. All participants were informed about the purpose of the research, the procedures, and their rights. We ensured strict confidentiality of all personal information and responses, and no data were linked to any identifiable personal information.

### 4.2 Procedure

In this experimental setup, we implemented four wearable design tasks that encompassed the most crucial areas of wearable technology [33]. These domains included fitness and wellness, entertainment, healthcare and medical, and industrial and =military. To ensure fairness in the prompt phrasing, we assigned four wearable design challenges to the four groups and conducted experiments by using the same task input for each group. Table 1 presents the tasks in detail.

Table 1: DETAILS OF CONCEPTUAL DESIGN TASKS

| Filed | Task 1: Fitness and Wellness | Task 2: Infotainment | Task 3: Healthcare and Medical | Task 4: Industrial and Military |
| --- | --- | --- | --- | --- |
| background | The intense consumer interest in personal health data has fueled the demand for smart fitness trackers, as users aim to improve their daily exercise and nutrition through the use of real-time data. | As the smartphone market becomes saturated and wearable device technology rapidly advances, glasses, watches, rings, pins, and even smaller ornaments could all become carriers for smart information generation and processing functions. | As the population ages, nursing homes are coming under increasing pressure to establish more efficient ways to monitor and care for the elderly. | In the field of construction machinery, remote control has become the focus. Increasingly more enterprises and customers need this convenient and safe method of interaction to control construction machinery and enable remote operation. |
| target_audience | Fitness enthusiasts, health-conscious consumers, and sports coaches. | Everyone | Nursing home administrators, nursing staff, elderly individuals, and their families. | Construction machinery operators, construction machinery design and manufacturing companies, and other construction machinery demanding units. |
| target_problems | Stay healthy and improve physical fitness. | Wearable personal mobile terminals that are more portable, smarter, and aligned with intuitive interaction logic. | Long-term health condition monitoring and early disease warning. | Improving the working environment for construction machinery operators, enhancing work efficiency, and achieving remote real-time control of construction machinery. |

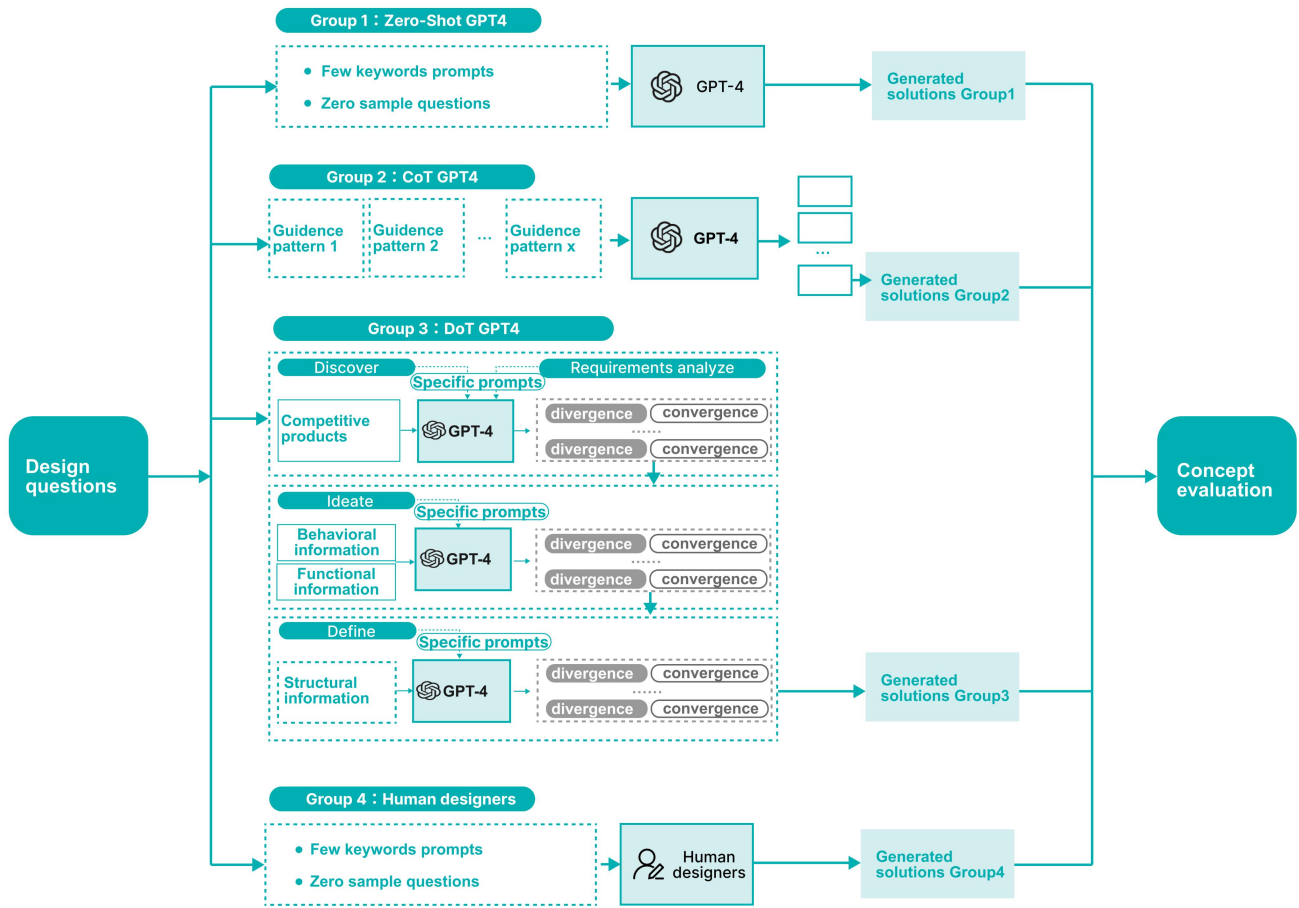

Fig.5: Settings for the experimental groups

### 4.3 Evaluation Metrics

Previous reports have extensively discussed the creative evaluation of design concepts, primarily focusing on the outcomes of these concepts [34, 35]. The consensual assessment technique (CAT) is the most widely recognised and used evaluation metric in this field. It leverages the diverse perspectives and knowledge of experts by utilising group consensus to enable a generally unbiased evaluation of creativity across various works. Alternatively, Shah et al. refined the following four dimensions to enable evaluation of the effectiveness and creativity of design concepts: quantity, novelty, quality, and variety [36, 37, 38]. Similarly, Xu et al. provided a set of design evaluation criteria that included two dimensions related to design quality, i.e., rationality and integrity. These dimensions facilitate detailed evaluation and analysis [39].

The novelty metric assesses whether design concepts offer unique solutions or perspectives that are essential for expanding the design space by exploring solutions that are not initially apparent. The variety metric measures the breadth of the solution space explored during the ideation process, emphasising the importance of examining multiple perspectives and creative restructuring problems. The quality metric focuses on the practicality and feasibility of ideas, focusing on the extent to which design solutions can meet established needs and factors such as cost and sustainability are reasonably considered. Furthermore, the total number of generated ideas supports the notion that generating more ideas increases the likelihood of producing superior solutions. Considering this, we established objective measures to evaluate the effectiveness of ideation, which has been widely adopted in previous research. Thus, our study applied these metrics. However, because we required designers and the GPT to provide a set number of design concepts for each design problem during the testing process, we opted to evaluate the design plans across three dimensions: novelty, variety, and quality, as shown in Figure 6. We further divided the quality dimension into rationality and completeness subdimensions to allow a more precise comparison and evaluation of the design concepts.

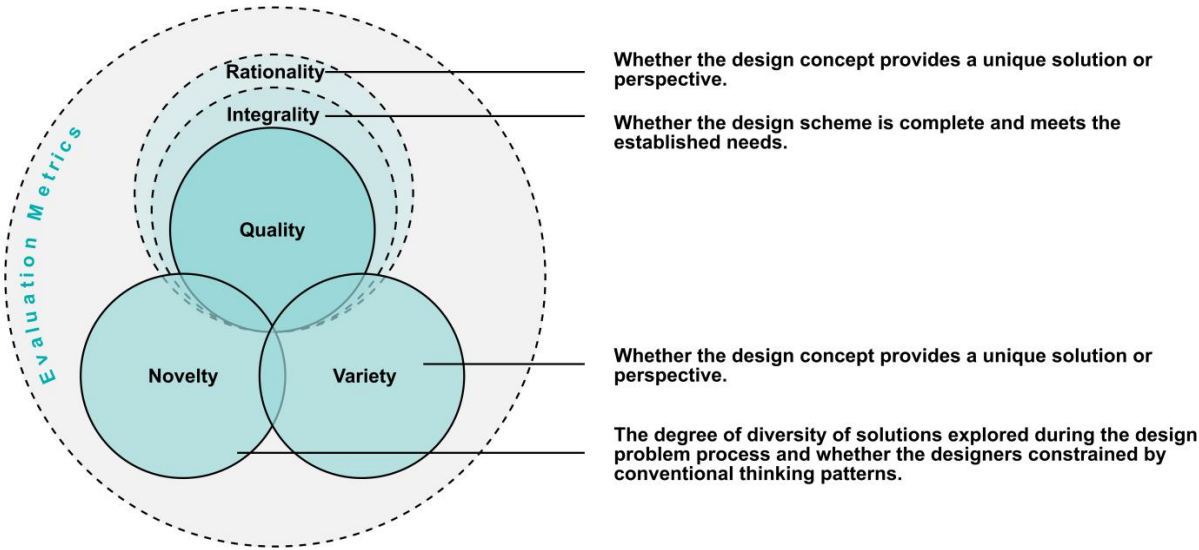

Fig.6: Evaluation metrics

### 4.4 Results

After the four groups completed the design tasks, we developed a questionnaire based on the design standards. Subsequently, we recruited six professional designers with more than five years of experience and a deep understanding of the concept of embodied cognition to serve as assessment experts. These experts independently evaluated the performance of the four designer types on the four design tasks.

Table 2: Consistency evaluation

| Metrics | Kendall's W |
|---|---|
| Rationality | 1.00 |
| Integrality | 0.90 |
| Novelty | 0.91 |
| Variety | 0.91 |

Using the CAT, we initially analysed data consistency. Given the extensive gradation of our rating scale, which provided evaluators with considerable flexibility, there were notable fluctuations in the numerical values. Consequently, we did not compute these results directly. Instead, we converted the rating data into a rank–order table and applied Kendall's law to assess consistency. As shown in Table 2, the final analysis revealed that all metrics reached the desired levels of correlation.

The final results are shown in Figure 7. Comparison of the total scores and individual indicators revealed a clear score ranking according to the established design scoring standards: IO < CoT < DoT. The difference between the human designers and DoTs was not significant ($p = 0.89$). Particularly, according to the experimental results, DoT enhanced the overall ability of the LLM to perform conceptual design tasks, yielded better effects than CoT, and closely matched human performance.

Regarding the rationality metric, IO was considered to e the least rational, whereas CoT effectively enhanced the rationality of design solutions, approaching the level of human designers, as in other tasks. With the support of existing case data from the wearables database, DoT received the highest score for rationality. In the completeness analysis, the rankings were as follows: IO < CoT < Human < DoT. DoT provided detailed information on the functionality, behaviour, and structure at each stage of the design process. Moreover, it was able to obtain specific and rational technical solutions through queries in the design database, which the other three methods could not achieve. Thus, DoT is better in terms of completeness. Regarding innovation, IO and CoT scored similarly, and their box plot structures revealed similar distributions of data. Although DoT ranked above them, the level of innovation from human designers surpassed that of DoT. Diversity dimension mainly focuses on the designers' ability to explore the design space during the divergent phase, with the data indicating the following order: IO < CoT < DoT < Human.

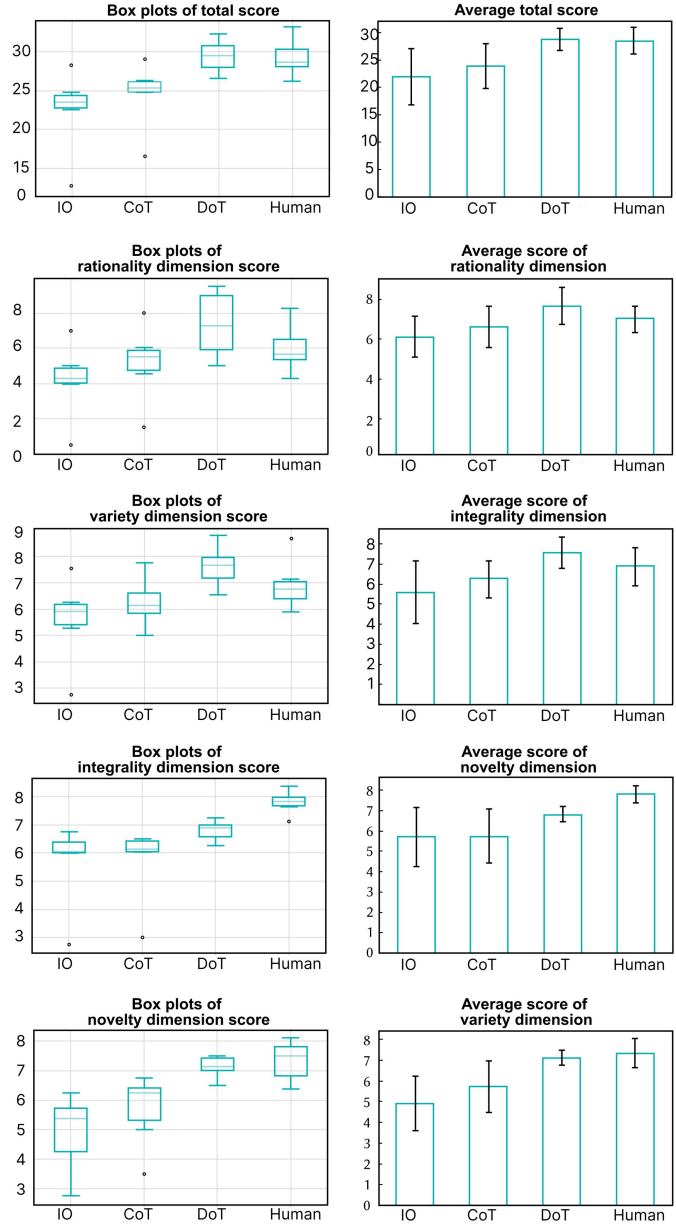

Fig.7: Evaluation results

We further compared the DoT and Human groups, finding that DoT significantly outperformed humans in terms of rationality and completeness, but was less effective in terms of innovation and diversity; these differences were statistically significant (p = 0.025, 0.039, 0.017 < 0.05, and p = 0.350 > 0.05, respectively). Notably, regarding diversity, although the statistical analysis did not reveal a significant difference between the DoT and humans (p = 0.350 > 0.05) at the selected level of significance, the observed difference between the two groups was relatively small. These subtle differences may be relevant for further research and analysis. Moreover, the small sample size may have limited the statistical power of the analysis. Thus, increasing the sample size could boost the statistical power, potentially leading to more conclusive results.

We also conducted additional enquiries for each design task, similar to the Turing test, wherein we asked the interviewed designers to assess whether the design solutions were conceived by human designers. The probabilities were obtained as presented in Table 3.

Table 3: Probability Evaluation

| IO | CoT | DoT | Human |
|---|---|---|---|
| 50% | 41.6% | 62.5% | 70.8% |

During the additional interviews, we made an enquiry regarding the basis of each judge's decision. The judges mentioned that the responses from the IO group were the easiest to critique because of their tendency to merely expand on and associate ideas based on a given design problem. Although the IO group provided extensive textual descriptions, their overall expressions were considered to be empty and lacking practical significance without deep analysis or concrete solutions. One judge commented that "[t]here is a lot written in the design proposal, but it lacked any new output, merely reiterating the design requirements." In contrast, responses for the CoT group indicated higher levels of consistency and coherence. The technical, functional, and material solutions mostly resulted in good correlations, confirming good structure and logicality. However, the responses were still insufficient. A judge noted that "[t]he technical and functional solutions in this response were closely linked and focused on four points, but the actual content seemed to be manufactured to meet a standard structure rather than being substantively rich." Alternatively, determining whether the design solutions were produced by DoT or humans proved difficult for many judges, as both groups were considered to have more effectively addressed design challenges and provided viable solutions. Particularly, the DoT group employed a mature problem-solving process for design challenges and performed detailed analyses of the design requirements. These characteristics led some judges to misjudge them, thinking that these responses may have been from experienced human designers: "This solution seems like it may have been written by a human because it uses an existing framework of thought or sequence of thought, although somewhat mechanised. It resembles human thinking and expression habits, such as using analogies with features of existing products when describing specific functional points and conducting requirement analyses to pinpoint the target user group."

During the interviews, we invited judges to individually assess whether each design solution incorporated the concept of embodied cognition. The judges generally believed that the solutions provided by the IO and CoT groups only considered embodied cognition in terms of multimodal feedback and cognitive load management, interpreting embodiment primarily by using the body's multimodal information processing channels and designing more intuitive and user-friendly interfaces. In contrast, the DoT group demonstrated a better understanding of the embodied concepts. Their design solutions appropriately integrated and applied the body's sensory systems to specific situations and designed more embodied interaction methods that aligned device operations with natural human behaviour and perception. This may be attributed to our database providing the DoT with extensive knowledge regarding how the functionality and structure of wearable products can optimise and affect human bodily perception and behavioural responses. For example, one DoT proposal mentioned the following: "By analysing user behavioural patterns and environmental interactions, the [artificial intelligence] butler can predict user needs and respond in advance, combining voice output and tactile feedback (such as vibration) to provide necessary guidance and feedback when it is inconvenient to look at the screen." Human designers' understanding of embodied cognition is deeper; for instance, one human designer mentioned utilising and transforming the embodied experience of experts in the field of engineering machinery: "Experts accumulate a significant amount of embodied experience during the regular operation of engineering machinery, which is valuable. Therefore, capturing the typical operational behaviours of experts can be transformed into intelligent guidance and support tips for smart gloves, thereby enhancing the accuracy of user operations."

## 5 discussion

We have developed and evaluated a DoT framework. Through comparative experiments, we demonstrated that the DoT framework effectively improved the performance of LLMs on conceptual design tasks.

Previous studies have highlighted the potential of LLMs on generative conceptual design tasks [3, 4, 5]. Ma et al. demonstrated that, on average, LLMs generate more feasible and useful designs than crowdsourced solutions, although the latter tend to offer more novelty [4]. This finding aligns with our results, which indicate that LLMs excel in integrating and applying diverse knowledge, which allows them to synthesise a wide array of information. This capability also allows LLMs to consider the technical, material, and functional aspects of design solutions from various interdisciplinary perspectives, thereby maximising their feasibility and practicality. Ma et al. also explored prompt engineering by using zero-shot and few-shot learning methods to enhance the quality of LLM responses, demonstrating that LLMs are highly influenced by the prompts they receive [4]. Implementing an effective prompt strategy significantly enhances the performance of LLM-generated design solutions. In our study, we carefully managed prompts by applying structured prompts at specific stages to guide the LLMs to complete the predefined design steps. Additionally, we integrated the database information into prompts, allowing LLMs to leverage existing case knowledge. However, we observed that LLMs tend to rely on pre-existing data and common patterns of thinking, which can limit their ability to generate truly novel ideas or "think outside the box." Therefore, establishing an appropriate strategy to co-optimise the problem-solving innovation capabilities of LLMs and rationality of conceptual design solutions remains an unresolved challenge. Yao et al. proposed the ToT method, which allows LLMs to use topological structures to simulate human problem-solving processes [24]. Similarly, DoT adopts a tree-like structure similar to ToT, but it is extended in our study by integrating predefined design steps and support from a design database, which contribute to enhance its ability to solve design problems. Wang et al. [5] introduced a methodology to address the challenges of transparency and controllability in LLMs by structuring the design process into decomposed tasks focused on functional, behavioural, and structural reasoning. They noted that current LLM-based generative design approaches primarily rely on direct generation, with no theoretical guidance or comprehensive understanding of the design requirements. This insight has inspired further research to leverage existing design theories and models to enhance LLM capabilities. Building on the incorporation of embodied cognition, we encoded the content of the wearable database according to the basic structure of the FBS model and input it into the LLM in stages. This approach allows the LLM to be supported by the foundational mapping relationships of the FBS system when analysing and generating specific design solutions. Additionally, design thinking, an essential theoretical and model foundation for problem solving in the design field, has not been adequately utilised. Therefore, we introduced the DoT framework by integrating design-thinking models and LLM task-processing capabilities. This framework addresses the existing gap and offers a potential systematic approach to using LLMs to solve design problems. In addition to applying predefined design tasks based on design-thinking models, we structured the application of divergent and convergent thinking in the LLM's design solution generation process, allowing for more effective and systematic creation and selection of innovative design concepts. To further enhance LLM controllability during the solution generation process, Wang et al. fine-tuned the model by using a precollected design knowledge dataset [5], allowing it to effectively utilise domain-specific knowledge for more relevant and innovative solutions. Similarly, we have incorporated extensive case studies into the design process through the use of enhanced RAG technology, broadening the knowledge and design themes accessible to LLMs.

Overall, the results of this study show that the DoT group scored slightly higher than the human designer group in terms of completeness and rationality. Based on interviews with scoring experts, we concluded that LLMs have a notable advantage over human designers in terms of ability to process and analyse large volumes of text data and integrate interdisciplinary knowledge, which has been widely recognised. In wearable design, the flexible application of interdisciplinary knowledge is crucial because it provides comprehensive support for solving design problems in various dimensions. In contrast, human designers with a single background may lack the specialised knowledge required to cover the entire design process. In this study, LLMs enhanced with existing knowledge and rules from a wearable design database were found to more effectively align FBS relationships, reduce the frequency of biases and fabrications, and demonstrate relatively superior performance in terms of rationality and completeness of solutions. Additionally, through iterative cycles of divergent and convergent thinking, LLMs can explore more design possibilities and quickly gather, filter, and match suitable information.

Nonetheless, the integration of the DoT framework into LLMs has some limitations. Regarding innovation, the DoT group scored lower than human designers. Although LLMs trained on wearable design case data can gain a solid understanding of existing cases and rules, this learning process also constrains their ability to innovate in problem solving. Creative thinking and imagination are crucial for design innovation; human designers can think abstractly and innovate when faced with the same design problems and case data, break existing frameworks, and generate entirely new ideas and concepts. This capacity for abstract thinking and innovation makes it challenging for LLMs to emulate them. Additionally, human designers possess the unique ability to empathise with and understand others' emotions and needs, often based on personal experiences and a deep understanding of the target user group. This emotional consideration cannot by replicated by current LLMs. Design is a flexible and complex iterative process that involves not only the application of explicit knowledge, but also the flexible utilisation of a designer's tacit knowledge and skills.

Designers often rely on their experience and intuition to make judgments based on deep insights and understanding accumulated through past design practices and social interactions. However, the informal and individualised nature of tacit knowledge makes it difficult for LLMs to fully convert it into symbolic knowledge for learning and understanding. Thus, a designer's tacit knowledge and skills are challenging for LLMs to acquire and replicate.

# 6 CONCLUSIONS

Here, we introduced and evaluated the DoT framework, a novel approach designed to enhance LLM capability in the interdisciplinary field of wearable design. Using a comparative experimental setup, we compared the performances of human designers and LLMs, which were enhanced through various techniques across several wearable design tasks. The findings clearly indicate the superior design problem solving capability of the DoT framework, demonstrating a level of comprehension that significantly surpasses those of conventional methods, such as direct IO and CoT frameworks. However, significant gaps remain in their ability to rival human designers in key areas such as innovation, abstract reasoning, and empathy.

Future research on applying LLMs to solve design problems should focus on enhancing their autonomy and flexibility by equipping them with a broader set of design methods, deeper design knowledge, and enhanced behavioural capabilities. Moreover, future studies should explore the collaborative dynamics between human designers and LLM agents, catalysing new forms of design and fostering a dynamic, innovative design environment.

## ACKNOWLEDGMENTS

The authors want to thank all the participants to the study for their availability.

## FUNDING

This work was supported by the National Key Research and Development Program of China [Grant Number 2021YFF0900602].

## DATA AVAILABILITY

The wearable database employed in this study is not publicly available due to the nature of the research. However, experimental data analyses that support the findings of this study are available from the corresponding author upon reasonable request.